\documentstyle[aps,preprint,psfig]{revtex}
\begin{document}
\draft

\title{Quantum Phase Transitions for Bosons in One Dimension}
\author{Reinhard Baltin\cite{baltin} and Karl-Heinz Wagenblast}
\address{Institut f\"ur Theoretische Festk\"orperphysik,
         Universit\"at Karlsruhe, D-76128 Karlsruhe, Germany}
\date{\today}
\maketitle
\begin{abstract}
  We study the ground state phase diagram and the critical properties of
  interacting Bosons in one dimension by means of a quantum Monte
  Carlo technique.  The direct experimental realization is a chain 
  of Josephson junctions.  For finite-range interactions we find a 
  novel intermediate phase which shows neither solid order nor superfluidity.
  We determine the location of this phase and study the critical 
  behaviour of the various transitions.  
  For on-site interaction only, we map out the phase diagram as a function
  of the hopping strength and the chemical potential.  
\end{abstract}

One dimensional systems attracted strong interest over the past years.
The major work focused on Fermionic models where a breakdown of
Landau's Fermi-liquid theory opens an avenue for new investigations.
Bosonic models deserve our attention for several reasons: The close
analogy between Fermions and Bosons in 1D, the study of
enhanced quantum fluctuations in low dimensions, and the possibility
of experimental investigation.

A good candidate for experimental studies is $^4$He, which has been
investigated in the bulk and on surfaces~\cite{bennemann}.  Another
class are superconductors.  The phase of the superconducting
order-parameter is a Bosonic quantum variable.  Nanofabricated
circuits, especially Josephson junction networks show these quantum
effects.  Mott transitions have been observed in two
dimensions~\cite{arrays} and recently also in one-dimensional
systems~\cite{oudenaarden}.  The low-energy Hamiltonian of a Josephson
chain coincides with that of a Luttinger liquid~\cite{fazio}.

The interplay between the interactions and the hopping gives rise to
various ground states which are separated by quantum phase
transitions.  A superconductor--insulator transition separates a
Mott-insulating phase with localized particles from a phase with
delocalized particles and superfluid response.  The critical value of
the hopping strength depends on the chemical potential.  The inclusion
of finite-range interactions enriches the picture, opening the
possibility for various commensurate Mott-insulating phases with a
solid-type ordering of the particles.
Mean-field studies \cite{liu} and numerical investigations of
two-dimensional systems \cite{otterlo,batrouni2,roddick} showed that the
transition from a Mott-insulating solid phase to the superfluid phase
splits into two separate transitions with intermediate supersolid
phases.  In these supersolid phases superfluidity and long-range solid
order coexist.  No such supersolid phase has been found in 1D
\cite{batrouni1}.

We study the quantum-phase model in 1D by means of a quantum Monte
Carlo method.  For on-site interactions we determine the phase diagram
with lobe-shaped Mott-insulating phases in good agreement with a
$t/U$-expansion of Freericks and Monien \cite{freericks}, as shown in
Fig.~1$a)$.  With on-site and nearest-neighbour
interactions the transition between solid phases, where the
site-occupancy alternates between $N$ and $N+1$, and superconducting
phase splits into two separate transitions as displayed by the scaling
plot in Fig.~2.  This gives rise to a novel {\em intermediate} phase
which shows neither superfluidity, nor a solid order.  We
determine the location of this phase and discuss its properties.
We extract the critical behaviour of 
the transitions in agreement with predictions of order-parameter
descriptions for the onset of superfluidity \cite{doniach,fwgf}. 

The quantum-phase model describes Bosons on a lattice.  Its Hamiltonian 
is given by
\begin{equation}
  H=\sum_{i}\left[\frac{U_0}{2} n_i^2 + U_1 n_i n_{i+1} 
    - \mu n_i - t \cos (\phi_{i+1}-\phi_{i})\right].
\label{ham}
\end{equation}
Two key ingredients are incorporated in this model: The interaction of
the Bosons (on-site $U_0$ and nearest-neighbours $U_1$), and the
hopping $t$.  The chemical potential $\mu$ tunes the number of
particles on the lattice.  Phase $\phi_i$ and number $n_i$ of this
model are non-commuting operators, i.e.\ $[n_i,\phi_j]=i\delta_{ij}$.
The variable $i$ labels the lattice sites.  The essential physics is
thus dominated by the competition between phase-coherence and solid
order.  This gives rise to quantum fluctuations and quantum
phase-transitions at zero temperature.  The properties of this model
are periodic in $\mu$ with a period of $U_0+2U_1$.  The quantum-phase
model directly represents a chain of Josephson junctions with $n_i$
being the excess number of Cooper pairs on site $i$, the Coulomb
interactions $U$, and the Josephson coupling $t$.  
A gate voltage can be applied to tune the chemical potential \cite{otterlo}.
The quantum-phase model is equivalent to
the Bose-Hubbard model in the limit of a large number of Bosons per
site \cite{otterlo}.

We use the mapping of a $d$-dimensional quantum model onto a
$d$+$1$-dimensional classical model to study the Hamiltonian of
Eq.~(\ref{ham}).  The 1D quantum-phase model has a
current-loop representation in 1+1 dimensions with the partition
function \cite{sorensen,otterlo}
\begin{eqnarray}
  Z=\sum_{\{J^{\tau},J^x\}}
  \delta(\partial_{\tau} J^{\tau}+\partial_x J^x)
  \exp\bigg\{-K\sum_{i,\tau}\Big[(J^\tau_{i,\tau})^2
  +\frac{U_1}{U_0}J^\tau_{i,\tau}J^\tau_{i+1,\tau}
  -\frac{\mu}{U_0}J^\tau_{i,\tau}+(J^x_{i,\tau})^2\Big]\bigg\}\;.
  \label{part}
\end{eqnarray}
A discrete two-component current $(J^x,J^\tau)$ flows on a
(1+1)-dimensional space-time lattice.  The constraint $\partial_{\tau}
J^{\tau}+\partial_x J^x=0$ allows only divergence-free current loops.
The imaginary time is discretized with the time-spacing $\epsilon$. 
The mapping is explained in detail in
Ref.~\onlinecite{otterlo}.  The effective coupling constant $K$
plays the role of the inverse temperature of the classical model and
is given by $K=\epsilon U_0/2=\ln\left[I_0(\epsilon t)/I_1(\epsilon
t)\right]$, where $I_0$, $I_1$ are the modified Bessel functions.
The time spacing $\epsilon$ is chosen to be of the order of the
inverse Josephson plasma-frequency, $\epsilon\approx 1/\sqrt{tU_0}$,
such that the couplings in $x,\tau$-direction are isotropic.  The time
component of the current $J_i^{\tau}$ can be identified as the number
of excess Bosons on site $i$.

The key quantities in our study are the superfluid stiffness
$\rho_0$ and the structure factor $S(k)$.  The former measures
superfluid correlations in the system.  The latter indicates whether
the particles are arranged periodically, i.e.\ whether the system is
in a solid ground state.  Both quantities can be expressed in terms of
the currents $J$ on a lattice of size $L\times
L_{\tau}$~\cite{sorensen,otterlo},
\begin{eqnarray}
  \label{str}
  \rho_0&=&\frac{1}{L L_\tau}
  \sum_{i,\tau} \langle J_{i,\tau}^x J_{0,0}^x \rangle,\\
  S(k)&=&\frac{1}{L L_\tau}
  \sum_{i,\tau} \langle J_{i,\tau}^{\tau} J_{0,\tau}^{\tau}
  \rangle\exp\{ikr_i\}.
\end{eqnarray}
Nearest-neighbour interactions give rise to solid phases with a finite
$\pi$-component of the structure factor $S_{\pi}=S(k=\pi)$.

An order-parameter description \cite{fwgf} for the onset of
superfluidity predicts a dynamical critical exponent $z=1$ in
the particle-hole symmetric case at the tips of the lobes in the
$t$-$\mu$ plane and a Kosterlitz-Thouless (KT) transition.  For broken
particle-hole symmetry away from these symmetry lines, $z=2$ and a
power law critical behaviour follows from these considerations.  
A Ginzburg-Landau description for the
transition to a finite $S_\pi$ was developed in
Ref.~\onlinecite{balents}.  In the absence of a superfluid background
this action implies the dynamical critical exponent $z=1$.

The simulation is confined to rather small system sizes, and we use
finite-size scaling for determining critical properties of our model.
For power-law critical behaviour we use the scaling
Ansatz~\cite{otterlo,batrouni1}
\begin{eqnarray}
  \label{scalrho}
  \rho_{0}&=&{L^{1-z}}\tilde{\rho}(L^{1/\nu}(t-t_c),L_{\tau}/L^{z}),
\end{eqnarray}
\begin{eqnarray}
  \label{scalspi}
  S_{\pi}&=&L^{-2\beta/\nu}\tilde{S}(L^{1/\nu}(t-t_c),L_{\tau}/L^{z}),
\end{eqnarray}
where $\beta$ is the critical exponent of the order parameter, $\nu$
is the critical exponent for the correlation length, and
$\tilde{\rho}, \tilde{S}$ are scaling functions.

A KT critical point can be determined using the jump of the stiffness,
characteristic logarithmic corrections, and the behaviour of the
exponentially diverging correlation length \cite{weber}.

We also investigated the possibility of a first order transition by
studying the energy histogram.  We find no evidence for a first order
transition in the phase diagram.

We simulate the model of Eq.~(\ref{part}) with periodic boundary
conditions using the Metropolis algorithm.  The current loops can be
divided into two classes.  Local loops represent a current around a
single plaquette on the lattice.  Global loops describe a net current
through the whole system.  In each Monte Carlo sweep we try to create
a local loop on each lattice site and global loops throughout the
lattice.  Each Monte Carlo run for one data point consists of $2\cdot
10^5 - 2\cdot 10^6$ sweeps for equilibration and $10^6-10^7$ sweeps
for measurement, depending on the lattice size and the coupling $t/U_0$.

We now present our numerical data, first for on-site interaction
($U_1=0$).  For non-integer values of $\mu/U_0$ particle-hole symmetry
is broken and according to Eq.~(\ref{scalrho}) the scaled data
$L\rho_0$ vs.\ $t$ for different lattice sizes should cross at the
critical coupling $t_{c}$ provided $L^2/L_{\tau}$ is kept constant.
Fig.~1$b)$ shows the scaled superfluid stiffness vs.\ the coupling for
$\mu/U_0=0.3$.  The curves for different lattices cross at
$t_c/U_0=0.207\pm0.003$.  The critical exponent $\nu$ is fitted such
that plots of $L\rho_0$ vs.  $L^{1/\nu}(t-t_c)$ collapse onto one
curve, the scaling function $\tilde{\rho}$, as shown in the inset of
Fig.~1$b)$.  We find $\nu=0.6\pm0.1$ which agrees with results from
\cite{fwgf}.
At $\mu/U_0=0.4$ the critical coupling is $t_c/U_0=0.10\pm0.005$.  For
$\mu/U_0=0.2$ we find $t_c/U_0=0.325\pm0.01$.  The critical exponent
$\nu$ is independent of the chemical potential in the range
$0.2\leq\mu/U_0\leq 0.4$.  At integer values of $\mu/U_0$, the system
has particle-hole symmetry.  Here we use the universal jump of the
superfluid stiffness characteristic for a KT transition to determine
the transition point~\cite{weber}.  From the simulations with
quadratic lattices ($L,L_{\tau}\le24$) we obtain $t_c/U_0=0.83\pm0.07$.

The resulting phase diagram consists of Mott-insulating lobes 
with fixed integer density in the $t$-$\mu$ plane, see Fig.~1$a)$.  The
cusp-like shape of the lobes is also found in the phase diagram of the
related 1d Bose-Hubbard model analyzed in Ref.~\onlinecite{batrouni1}.
In the vicinity of the symmetry lines at integer values of $\mu/U_0$
we find deviations from the predicted finite-size scaling of the
superfluid stiffness of Eq.~(\ref{scalrho}).  We argue that this is
due to the vicinity of the KT transition which introduces another
length scale in the scaling relation.  Asymptotically for large
systems we expect to recover a scaling according to
Eq.~(\ref{scalrho}).  The solid lines in Fig.~1$a)$ are the phase
boundary obtained by a third order $t/U$-expansion for the
Bose-Hubbard model of Freericks and Monien~\cite{freericks} in the
limit of a large number of Bosons per site which is in good agreement
with our data.  In Ref.~\cite{freericks} it is argued that the
deviation near the tips of the lobes is due to the KT transition whose
physics cannot be described in perturbation theory of finite order.

When nearest-neighbour interaction is included the phase diagram is
richer, including solid, superfluid and novel intermediate phases.  We
find Mott-insulating lobes where the site occupancy alternates
periodically between $N$ and $N+1$, i.e.\ solid order with a finite
structure factor $S_\pi$, a superfluid phase with $\rho_0\neq 0$, and
a compressible intermediate phase with $S_\pi=0$ and $\rho_0=0$.  We
focus in our studies on the case of broken particle-hole
symmetry where we can use the scaling relations of Eq.~(\ref{scalrho})
and Eq.~(\ref{scalspi}).  The dynamical exponent for the onset of
superfluidity is thus $z=2$.  For the transition to a finite $S_\pi$
we use $z=1$, and fit the ratio $2\beta/\nu$ such that the scaled data
cross in one point.  We estimate for $2\beta/\nu=0.12\pm0.02$.  Fig.~2
shows the scaled data of $\rho_0$ and $S_{\pi}$ vs.\ coupling for
$U_1/U_0=0.4$ and $\mu/(U_0+2U_1)=0.6$.  We obtain for the superfluid
transition $t_c/U_0=0.130 \pm 0.006$ and for the transition to the
solid phase $t_c/U_0=0.112 \pm 0.005$.  In between there is an
intermediate phase in which the system neither is superfluid nor solid.
We address this lack of order at $T=0$ to the enhanced
fluctuations in 1D.  There are corrections to the scaling relation of
the superfluid stiffness, since the curves in Fig.~2 do not exactly
cross in one point.  A detailed investigation of these corrections,
which we address again to the vicinity of the KT transition, is beyond
our numerical resolution.

The absence of both, superfluidity and solid order, may imply the
existence of a normal phase of the Bosons with metallic response.  A
study of the superfluid stiffness as a function of the Matsubara
frequencies and an analytic continuation can yield insight into the
response properties \cite{sorensen}.  Within the accuracy
of our results we are not able to prove or disprove the existence of a
finite d.c.\ conductivity \cite{diplom}.  There are general objections
to a normal phase for Bosons at zero
temperature by Leggett~\cite{leggett}.  The arguments are based on the
absence of nodes of the many body ground state wave function for
Bosons in the continuum.  These arguments are not directly applicable
to our case as we study a lattice model.  Furthermore, Leggett himself
argued that the existence of a normal ground state of Bosons cannot be
ruled out completely in 1D.  The question of the response of the
intermediate phase remains open and subject to further studies.  
A very recent work predicts the existence of a repulsive Luttinger 
liquid intervening the solid and superfluid phase \cite{larkin}.  

In conclusion we determined the phase diagram of the quantum-phase
model in 1D by means of a quantum Monte Carlo method.  For repulsive
on-site interaction we mapped out the phase diagram as a function of
the hopping strength and the chemical potential.  Finite-range
interactions give rise to new phases.  The insulating phases with
half-integer filling have solid order with a unit cell of two lattice
spacings.  Our simulations show an intermediate phase where both,
solid and superfluid order are destroyed by the strong quantum
fluctuations in 1D.

We would like to thank Rosario Fazio, Anne van Otterlo, Gerd Sch\"on,
and Gergely T.~Zimanyi for stimulating discussions.  This work is
within the SFB195 of the ``Deutsche Forschungsgemeinschaft''.

\newpage
\begin{figure}[b]
  \vspace*{2cm}
  \begin{center}
    \leavevmode
    \parbox{0.5\textwidth}
    {\psfig{file=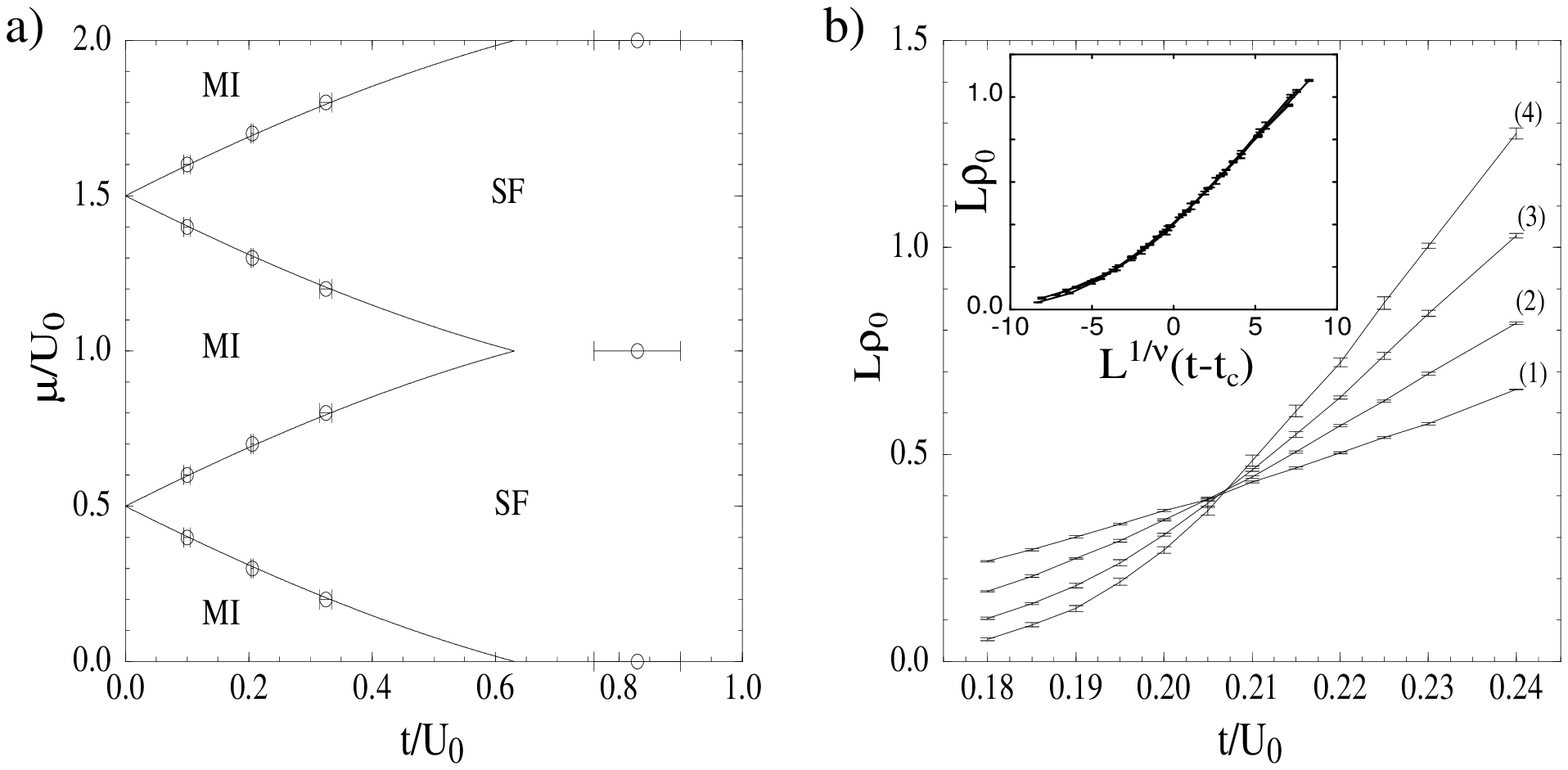,width=\textwidth,height=0.5\textwidth,angle=-90}}
\end{center}
  \vspace*{2cm}
  \caption[]{
    Results for on-site interaction. $a)$ Phase diagram.  The phase
    boundary separates the Mott-insulating (MI) phase from the
    superfluid phase (SF).  The symbols are our Monte Carlo results,
    the solid line is the result of a third order $t/U$-expansion from
    Ref.~\protect{\onlinecite{freericks}}.  
    $b)$ Scaled data for the superfluid stiffness at $\mu/U_0=0.3$.
    The intersection of the curves gives the transition point at
    $t_c=0.207\,U_0$.  Array sizes $L\times L_{\tau}$: (1) $6\times
    9$, (2) $8\times 16$, (3) $10\times 25$, (4) $12\times 36$.  The
    inset shows the scaling of the data to a single function according
    to Eq.~(\protect{\ref{scalrho}}), with $\nu=0.6$.  }
\end{figure}

\newpage
\begin{figure}[b]
  \vspace*{2cm}
  \begin{center}
    \leavevmode
    \parbox{0.5\textwidth}
{\psfig{file=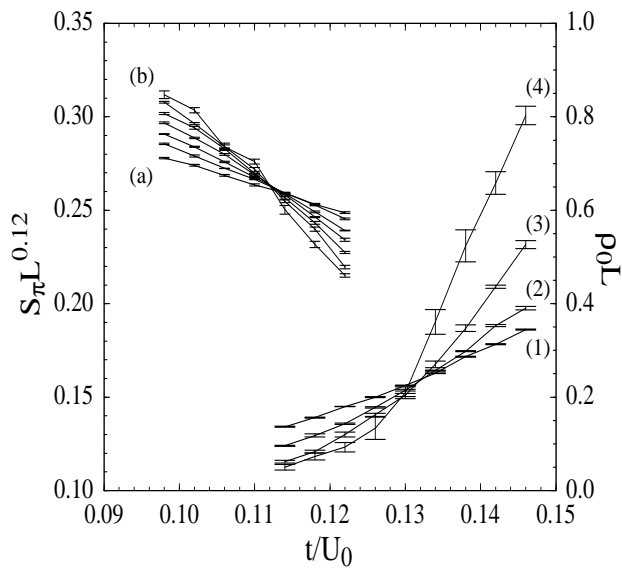,width=0.5\textwidth,height=0.5\textwidth,angle=0}}
  \end{center}
  \vspace*{2cm}
  \caption{
    Scaled data for the structure factor (left) and the superfluid
    stiffness (right) for $U_1/U_0=0.4$ and $\mu/(U_0+2U_1)=0.6$.  
    In the intermediate phase for $0.112<t/U_0<0.130$
    the system shows neither solid order nor superfluidity.
    Array sizes $L\times L_{\tau}$ for the superfluid stiffness: 
    (1) $6\times 9$, (2) $8\times 16$, (3) $10\times 25$, (4) $12\times 36$.
    For the structure factor: $L=L_\tau$ = 8 (a), 10, 12, 14, 16, 18, and 
    20 (b).}
\end{figure}

\end{document}